\documentclass[conference]{IEEEtran}
\ifCLASSINFOpdf
\else
\fi
\usepackage{cite}
\usepackage[cmex10]{amsmath}
\usepackage{multirow}
\usepackage{array}
\usepackage[lofdepth,lotdepth]{subfig}
\usepackage{lipsum}
\usepackage[pdftex]{graphicx}
\usepackage{mathtools}
\usepackage{cuted}
\usepackage{breqn}
\usepackage{graphicx}
\usepackage{gensymb}
\usepackage[english]{babel}
\usepackage[autostyle]{csquotes}
\usepackage[acronym,toc,shortcuts]{glossaries}
\usepackage{caption}
\usepackage{xcolor}

\makeglossaries
\newacronym{ACO-OFDM}{ACO-OFDM}{Asymmetrically Clipped Optical OFDM}
\newacronym{ADAS}{ADAS}{advanced driver-assistance systems}
\newacronym{APD}{APD}{avalanche photodiode}
\newacronym{ANN}{ANN}{artificial neural network}
\newacronym{AVP}{AVP}{autonomous valet parking}
\newacronym{AWGN}{AWGN}{Additive White Gaussian Noise}
\newacronym{BER}{BER}{bit-error-rate}
\newacronym{CAV}{CAV}{connected and autonomous vehicle}
\newacronym{CC}{CC}{Convolutional Code}
\newacronym{CDF}{CDF}{cumulative density function}
\newacronym{C-V2X}{C-V2X}{Cellular Vehicular to Everything Communication}
\newacronym{CLT}{CLT}{Central Limit Theorem}
\newacronym{CIR}{CIR}{Channel Impulse Response}
\newacronym{CFR}{CFR}{Channel Frequency Response}
\newacronym{CMOS}{CMOS}{Complementary Metal Oxide Semiconductor}
\newacronym{COTS}{COTS}{commercial off-the-shelf}
\newacronym{CSK}{CSK}{Color Shift Keying}
\newacronym{DC}{DC}{direct current}
\newacronym{DCO-OFDM}{DCO-OFDM}{Direct Current Biased Optical OFDM}
\newacronym{DLoS}{DLoS}{directed-line-of-sight}
\newacronym{DPIM}{DPIM}{Digital Pulse Interval Modulation}
\newacronym{DRL}{DRL}{Day Time Running Light}
\newacronym{DSRC}{DSRC}{Dedicated Short Range Communication}
\newacronym{EUB}{EUB}{effective usable bandwidth}
\newacronym{EMI}{EMI}{electromagnetic interference}
\newacronym{FEC}{FEC}{Forward Error Correction}
\newacronym{FLP}{FLP}{fast locking pattern}
\newacronym{FFT}{FFT}{Fast Fourier Transform}
\newacronym{FoV}{FoV}{field of view}
\newacronym{FPGA}{FPGA}{Field Programmable Gate Array}
\newacronym{GPSDO}{GPSDO}{Global Positioning System disciplined oscillator}
\newacronym{IF}{IF}{intermediate frequency}
\newacronym{IDFT}{IDFT}{inverse discrete Fourier transform}
\newacronym{IFFT}{IFFT}{inverse Fast Fourier Transform}
\newacronym{IM/DD}{IM/DD}{intensity modulation / direct detection}
\newacronym{IR}{IR}{infrared}
\newacronym{ISI}{ISI}{inter-symbol interference}
\newacronym{LSTM}{LSTM}{long-short-term-memory}
\newacronym{C-ITS}{C-ITS}{Cooperative Intelligent Transportation Systems}
\newacronym{LED}{LED}{Light Emitting Diode}
\newacronym{LNA}{LNA}{low noise amplifier}
\newacronym{LoS}{LoS}{line-of-sight}
\newacronym{LTE}{LTE}{Long Term Evolution}
\newacronym{MAC}{MAC}{medium access control}
\newacronym{MCS}{MCS}{modulation coding schemes}
\newacronym{MMSE}{MMSE}{minimized mean squared error}
\newacronym{MPC}{MPC}{multi-path component}
\newacronym{MIMO}{MIMO}{multiple input multiple output}
\newacronym{mmWave}{mmWave}{millimeter-wave}
\newacronym{NLoS}{NLoS}{non-line-of-sight}
\newacronym{NBV}{NBV}{nearby vehicle scenario}
\newacronym{OBU}{OBU}{on-board unit}
\newacronym{OOK}{OOK}{on-off keying}
\newacronym{OLoS}{OLoS}{obstructed line-of-sight}
\newacronym{OFDM}{OFDM}{orthogonal frequency division multiplexing}
\newacronym{QAM}{QAM}{Quadrature Amplitude Modulation}
\newacronym{PD}{PD}{photo-detector}
\newacronym{PDP}{PDP}{power delay profile}
\newacronym{PHR}{PHR}{physical header}
\newacronym{PHY}{PHY}{physical layer}
\newacronym{PMT}{PMT}{photo multiplier tube}
\newacronym{PN}{PN}{pseudorandom noise}
\newacronym{PPM}{PPM}{pulse position modulation}
\newacronym{PSDU}{PSDU}{physical service data unit}
\newacronym{PWM}{PWM}{pulse width modulation}
\newacronym{RF}{RF}{radio frequency}
\newacronym{RLL}{RLL}{Run-Length Limited}
\newacronym{RMS}{RMS}{root mean square}
\newacronym{RNN}{RNN}{recursive neural network}
\newacronym{RS}{RS}{Reed Solomon}
\newacronym{RSS}{RSS}{received signal strength}
\newacronym{RSU}{RSU}{road-side unit}
\newacronym{RU}{RU}{Receiver unit}
\newacronym{SDR}{SDR}{Software-Defined Radio}
\newacronym{SER}{SER}{Symbol Error Rate}
\newacronym{SHR}{SHR}{synchronization header}
\newacronym{SNR}{SNR}{signal to noise ratio}
\newacronym{TDL}{TDL}{tap delay line}
\newacronym{TDNN}{TDNN}{time delay neural network}
\newacronym{TIA}{TIA}{transimpedance amplifier}
\newacronym{TDP}{TDP}{topology dependent pattern}
\newacronym{TU}{TU}{transmitter unit}
\newacronym{U-OFDM}{U-OFDM}{Unipolar OFDM}
\newacronym{USRP}{USRP}{Universal Software Radio Peripheral}
\newacronym{VLC}{VLC}{visible light communication}
\newacronym{VVLC}{VVLC}{vehicular visible light communication}
\newacronym{VNA}{VNA}{vector network analyzer}
\newacronym{V2I}{V2I}{vehicle-to-infrastructure}
\newacronym{I2V}{I2V}{infrastructure-to-vehicle}
\newacronym{ITS}{ITS}{intelligent transportation systems}
\newacronym{V2V}{V2V}{vehicle-to-vehicle}
\newacronym{V2X}{V2X}{vehicle-to-everything}
\newacronym{V2P}{V2P}{vehicle-to-pedestrian}
\newacronym{V2LC}{V2LC}{vehicular visible light communications}
\newacronym{OWC}{OWC}{Optical Wireless Communications}
\newacronym{VPPM}{VPPM}{variable pulse position modulation}
\newacronym{LOS}{LOS}{line-of-sight}
\newacronym{PRBS}{PRBS}{pseudorandom binary sequence}
\newacronym{NI}{NI}{National Instruments}

\title{Vehicular Visible Light Communications for Automated Valet Parking}

\author{\IEEEauthorblockN{Bugra Turan$^1$, Ali Uyrus$^1$, Osman Nuri Koc$^2$, Emrah Kar$^2$, and 
Sinem Coleri $^1$} 
\IEEEauthorblockA{$^1$Department of Electrical and Electronics Engineering, Koc University, Sariyer, Istanbul, Turkey, 34450\\
$^2$Koc University Ford Otosan Automotive Technologies Laboratory (KUFOTAL), Sariyer, Istanbul, Turkey, 34450\\
E-mail: [bturan14, auyrus18, okoc, ekar, scoleri] @ku.edu.tr}}

\begin{document}

\maketitle
\begin{abstract}

\Ac{VLC} is a promising \ac{OWC} scheme which is demonstrated to provide secure, \ac{LoS}, and short distance \ac{V2V} and \ac{V2I} communications. Recently, automated driving applications, supported by \ac{V2I} links are proposed to increase the reliability of the autonomous vehicles. To this regard, we propose a \ac{VLC} based \ac{V2I} scheme to increase the \ac{V2I} communication redundancy of \ac{AVP} applications, through jam-free and location based characteristics of \ac{VLC}. In this paper, we demonstrate a novel architecture to support indoor parking garage online-map update with vehicle on-board data transmissions and location based map update dissemination through bi-directional \ac{VLC} communications. The proposed system yields error free \ac{LoS} transmissions with \ac{DCO-OFDM} up to 33 m transmitter - receiver distance enabling vehicle CAN Bus data, infrastructure camera video and LIDAR point cloud data sharing in an indoor parking garage. 
\end{abstract}


\IEEEpeerreviewmaketitle


\section{Introduction} 

\Ac{VLC} is proposed as a complementary technology to \ac{RF} based \ac{V2V} and \ac{V2I} communication systems, utilizing readily available \ac{LED} lights of vehicles and infrastructure as the transmitter with optical detectors as the receiver. Since \ac{VLC} offers \ac{RF} interference and jam-free secure \ac{LoS} communications, \ac{VLC} systems are foreseen to complement \ac{RF} based vehicular communication schemes for short distance, mission critical and precision oriented applications, such as platooning and \ac{AVP}. 

\ac{AVP} allows drivers to drop off and pick up their autonomous vehicle in front of a parking lot without taking care of parking. \ac{AVP} applications aim to provide numerous benefits, including refueling, cleaning and recharging of the vehicles, without human intervention \cite{vcharge2016, survey}.
Current \ac{AVP} implementations\footnote{https://www.bosch.com/stories/automated-valet-parking/} mainly rely on vehicle on-board perception sensors and pre-loaded offline maps to navigate where cellular connectivity is utilized to reserve the parking spot. 
However, the reliability issues of vision and LIDAR sensors is a concern, since vehicle vision sensors are challenged by the low illumination and highly reflective environment due to shiny road surface whereas LIDAR sensors suffer from the non-reflective surface and specular reflections. 

Since \ac{AVP} implementations solely based on vehicle perception sensors, vehicle lacks real time infrastructure sensor information 
regarding maneuvering vehicles and pedestrians beyond its perception sensor range. Moreover, the infrastructure lacks immediate vehicle information to guide the vehicles in the parking garage. Therefore, an online-map, constantly updated through sensor-equipped infrastructure and vehicle on-board information is key to enhance the reliability of the application. Furthermore, real-time online-map sharing with the relevant maneuvering vehicles helps to increase the navigation performance, with the prior knowledge of dynamic objects beyond sensor perception range.


In this paper, we present a VLC based \ac{V2I} communication scheme to increase the reliability of \ac{AVP}, and evaluate its \ac{BER} performance through real world measurements. The focus is on the integration of  \ac{VLC} based communications into to the indoor parking garage, which provides location based, need-to-know basis information to the vehicle and infrastructure to generate and share highly detailed online parking garage maps supporting \ac{AVP}. The proposed \ac{VLC} based scheme takes charge of sending location based vehicle on-board information to the infrastructure, and streaming up-to-date map information including pedestrian and maneuvering vehicle. The proposed scheme is foreseen to be a low cost, low complexity, and scalable communication solution to increase the reliability of \ac{AVP} applications.

\begin{figure}[h]
\centering
\includegraphics[clip,trim=0 0 0 0,width=\linewidth]{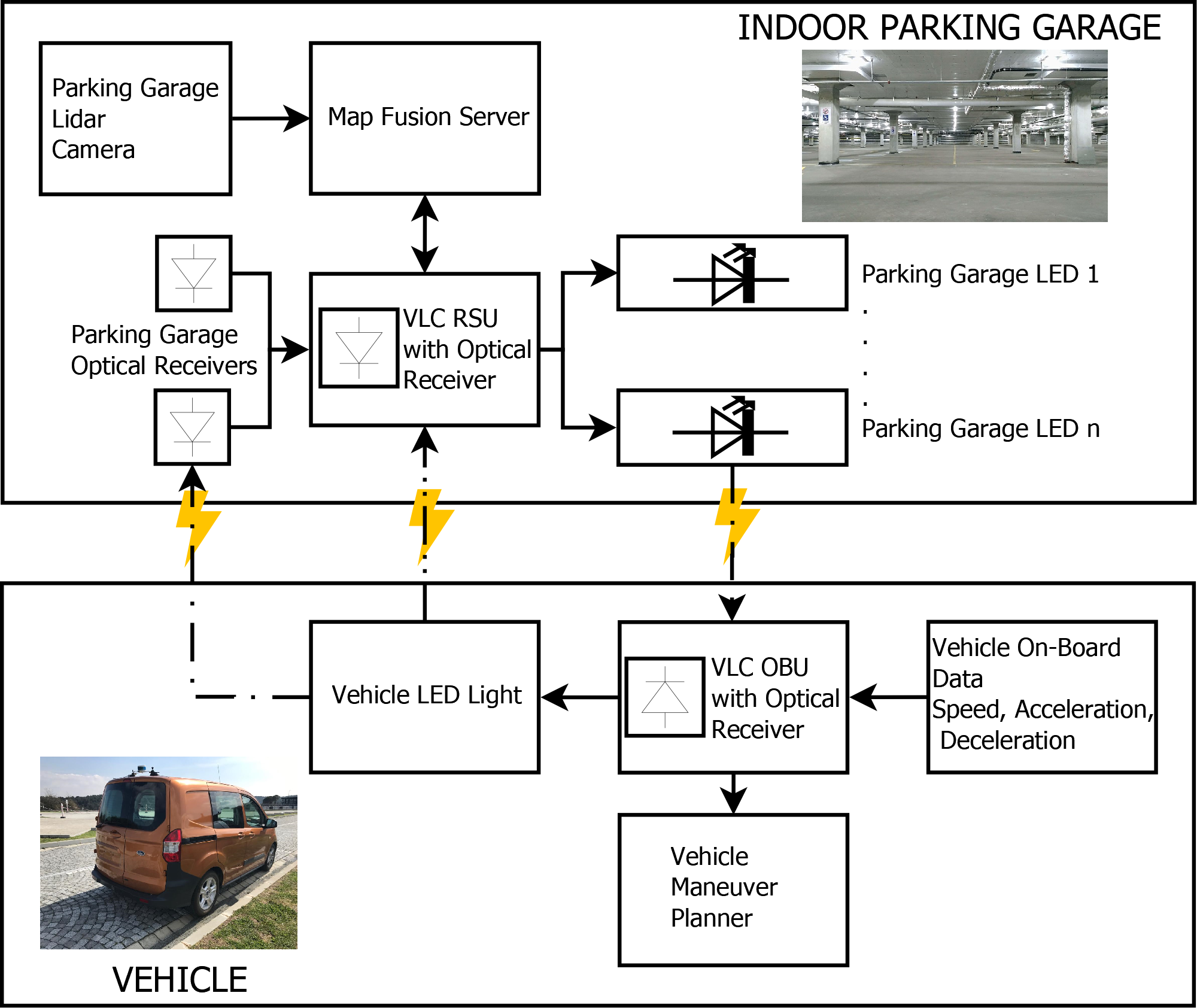}
\caption{\ac{VLC} based \ac{V2I} communications system architecture.}
\label{Fig1}
\end{figure}

\section{System Overview}

The \ac{VLC} based \ac{V2I} communications system is based on \ac{VLC}- \ac{OBU} , \ac{VLC}- \ac{RSU}, optical detectors and \glspl{LED}.  
The system architecture is depicted in Fig.\ref{Fig1}.

The vehicle transmits on-board data through \ac{VLC}-\ac{OBU} driven \ac{LED} lights. Moreover, the infrastructure information is captured and conveyed to the vehicle maneuver planner by the \ac{VLC}-\ac{OBU} of the vehicle. 


The online-map of the parking space is created with camera, LIDAR, and vehicle on-board information. The map server of the infrastructure is responsible to fuse and disseminate the online-map updates through infrastructure \ac{LED} lights driven by \ac{VLC}-\ac{RSU}. \ac{VLC}-\ac{RSU} also captures vehicle on-board information by optical receivers. 

\ac{VLC} provides a unique advantage of location based transmissions due to its \ac{LoS} characteristics. Therefore, only the relevant part of the online-map is transmitted from the infrastructure \glspl{LED}. On the other hand, even though vehicle continuously transmits information through its \ac{LED} lights, the optical receivers deployed on the infrastructure capture vehicle data only at certain locations, constrained by the detector's \ac{LoS} and \ac{FoV}. Therefore, the \ac{VLC} based \ac{V2I} communications enables location specific information exchange between vehicles and infrastructure, decreasing on-board computation needs, eliminating unnecessary information, while providing \ac{RF} interference and jam-free secure communications.

\section{System Description}

In this paper, we demonstrate a \ac{V2I} scheme, that enables information exchange with \ac{LED} lights and optical sensors of the vehicle and infrastructure. The vehicle transmits deceleration, acceleration, speed and brake status information and receives LIDAR point cloud map and video from the infrastructure with \ac{VLC}. On the other hand, optical detectors deployed in the parking garage receive vehicle data to aggregate the information into the LIDAR point cloud map. The constantly updated aggregated map is disseminated by infrastructure \ac{LED} lights to the vehicle. 

\begin{figure}[h]
\centering
\includegraphics[clip,trim=0 0 0 0,width=\linewidth]{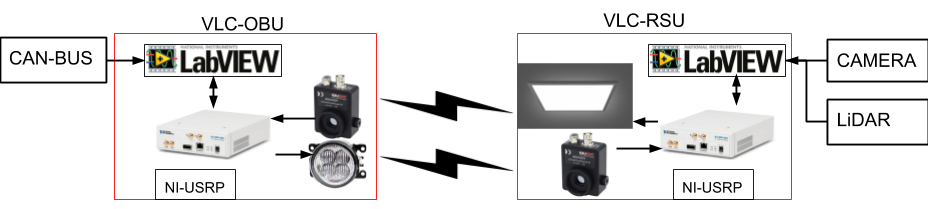}
\caption{\ac{VLC} based \ac{V2I} communications system architecture.}
\label{Fig.Demo}
\end{figure}

The system setup is depicted in Fig.\ref{Fig.Demo}. At the vehicle side \ac{VLC}-\ac{OBU} is responsible to convert vehicle CAN Bus information into \ac{DCO-OFDM} frames and drive vehicle \glspl{LED} with baseband \ac{DCO-OFDM} signals and required DC bias. Moreover, it incorporates an optical receiver to capture information from infrastructure \glspl{LED}. \ac{VLC}-\ac{OBU} is controlled with LabView software and equipped with a \ac{SDR}(NI-2920 with LFTX/RX Daughter board), \ac{LED} driver circuit, and an optical photodetector (Thorlabs PDA100A).

At the infrastructure, \ac{VLC}-\ac{RSU} converts LIDAR point cloud data into \ac{DCO-OFDM} frames, drives the infrastructure \glspl{LED} with base-band \ac{DCO-OFDM} signals and required DC bias. \ac{VLC}-\ac{RSU} incorporates the same hardware with \ac{VLC}-\ac{OBU}, where it is optimized to continuously stream map information instead of periodic vehicle on board information.

\begin{table}[h!]
\centering
\caption{System Setup Specifications}
\label{tab:OFE}
\begin{tabular}{lc}
\hline 
\textbf{Parameter} & \textbf{Value} \\ \hline
IQ Rate & $200$ kHz \\
Modulation & 4-QAM DCO-OFDM \\
$N_{FFT}$ & $64$\\
$N_{PILOT}$ & $4$ \\
Data Rate & $375$ kbps \\ \hline \hline
\end{tabular}
\end{table}

\section{Evaluation}
To determine the reliability of the considered \ac{VLC} based \ac{V2I} scheme, we evaluate the \ac{BER} performance of \ac{V2I} \ac{VLC} links with respect to varying transmitter-receiver distance and receiver angles to incorporate \ac{LoS} and \ac{DLoS} scenarios. System setup specifications are listed in Table \ref{tab:OFE}. For \ac{LoS} \ac{VLC} link of 39 $m$, $2.4\times10^{-5}$ \ac{BER} is achieved with 4.83 $dB$ \ac{SNR}, whereas $4\times10^{-5}$ \ac{BER} is obtained for \ac{DLoS} configuration with 20.98 $dB$ \ac{SNR}. For \ac{DLoS} the receiver is inclined towards the parking structure wall, where it captures distorted signal through reflection and scattering along with partial \ac{LoS} signals. Moreover, error free transmissions up to 33 $m$ \ac{VLC} link distance without any error correction coding is observed with a minimum \ac{SNR} value of 23.42 $dB$. Therefore, \ac{VLC} can be considered as a viable candidate to increase the redundancy of \ac{V2I} indoor parking garage links of \ac{AVP} applications.  

\section{Summary}
The \ac{VLC} based online-map update and vehicle data dissemination is presented as a candidate technology to increase the reliability of indoor parking garage applications including \ac{AVP}. Since, \ac{VLC} based \ac{V2I} schemes depend on spatial isolation, they are favorable for location based, low complexity, interference and congestion free \ac{V2I} communications. The proposed architecture is evaluated with real world implementations, where indoor parking garage is experienced to provide stable \ac{VLC} links including dynamic scenarios, as they provide ambient noise free channel for \ac{VLC}. Moreover, due to multi-path fading resiliency of \ac{VLC}, reflections from ground surface, walls, and nearby objects are observed to have limited adverse effects on the system performance. 
        


\bibliographystyle{ieeetr}
\bibliography{references}

\begin{thebibliography}{1}

\bibitem{vcharge2016}
U.~{Schwesinger} {\em et~al.}, ``Automated valet parking and charging for
  e-mobility,'' in {\em 2016 IEEE Intelligent Vehicles Symposium (IV)},
  pp.~157--164, 2016.

\bibitem{survey}
H.~Banzhaf, D.~Nienhüser, S.~Knoop, and J.~Zöllner, ``The future of parking:
  A survey on automated valet parking with an outlook on high density
  parking,'' pp.~1827--1834, 06 2017.

\end{thebibliography}
\nocite{*}

\end{document}